\def\be{\begin{equation}}
\def\ee{\end{equation}}
\def\ba{\begin{array}}

\def\ea{\end{array}}

\documentclass[12pt]{article}
\usepackage{amsfonts,amssymb,amsmath,graphicx}
\begin{document}
\parskip=3pt
\parindent=18pt
\baselineskip=20pt
\setcounter{page}{1}
\centerline{\large\bf Quantum Imaginarity-Mixedness Trade-off:}
\centerline{\large\bf Characterizing Maximally Imaginary Mixed States}
\vspace{6ex}
\centerline{{\sf Bin Chen,$^{\star}$}
\footnote{\sf Corresponding author: chenbin5134@163.com}
~~~ {\sf Shao-Ming Fei$^{\natural,\sharp}$}
}
\vspace{4ex}
\centerline
{\it $^\star$ School of Mathematical Sciences, Tianjin Normal University, Tianjin 300387, China}\par
\centerline
{\it $^\natural$ School of Mathematical Sciences, Capital Normal University, Beijing 100048, China}\par
\centerline
{\it $^\sharp$ Max-Planck-Institute for Mathematics in the Sciences, 04103 Leipzig, Germany}\par
\vspace{6.5ex}
\parindent=18pt
\parskip=5pt
\begin{center}
\begin{minipage}{5in}
\vspace{3ex}
\centerline{\large Abstract}
\vspace{4ex}

We investigate the trade-off relations between imaginarity and mixedness in arbitrary $d$-dimensional quantum systems. For given mixedness, a quantum state with maximum imaginarity is defined to be a ``maximally imaginary mixed state" (MIMS). By using the $l_{1}$ norm of imaginarity and the normalized linear entropy, we conclusively identify the MIMSs for both qubit and qutrit systems. For high-dimensional quantum systems, we present a comprehensive class of MIMSs, which also gives rise to complementarity relations between the $1$-norm of imaginarity and the $1$-norm of mixedness, as well as between the relative entropy of imaginarity and the von Neumann entropy. Furthermore, we examine the evolution of the trade-off relation for single-qubit states under four specific Markovian channels: bit flip channel, phase damping channel, depolarizing channel and amplitude damping channel.\\

\end{minipage}
\end{center}

\newpage

\section{Introduction}

Quantum systems subject to unavoidable environmental interactions which lead to quantum decoherence. To comprehensively assess how decoherence affects information processing, it is imperative to quantify the degree of state mixedness. In the work of Singh \emph{et al.} \cite{Singh}, a trade-off relation was established between the $l_{1}$ norm of coherence (a measure of quantum coherence) and the normalized linear entropy which quantifies the mixedness. Furthermore, they introduced the concept of maximally coherent mixed states (MCMSs) which have the maximum coherence for given mixedness. Since then such trade-offs involving various coherence quantifiers and mixedness measures have been explored \cite{Tao,chenmf,Sun}. Besides, investigating the trade-off relations between other quantum resource quantifiers and the extent of quantum state mixedness is also of significant interest.

As one of the most important quantum resources, the imaginarity plays a crucial role in both quantum mechanics and quantum information theory \cite{HG,Wu1,Wu2}. The rigorous framework for the resource theory of imaginarity was initially proposed by Hickey and Gour \cite{HG},  wherein the free states are defined as those given by real density matrices with respect to a fixed orthonormal basis. The corresponding free operations are described by $\Lambda(\cdot)=\sum_{j}K_{j}\cdot K_{j}^{\dag}$ with real Kraus operators $\{K_{j}\}_{j}$. Imaginarity has been quantified by imaginarity measures such as $1$-norm-based measure \cite{HG}, the robustness of imaginarity \cite{HG}, the geometric measure for pure states \cite{Wu1}, the relative entropy of imaginarity \cite{qip} and the $l_{1}$ norm of imaginarity \cite{sc}. These measures not only reveal the non-real components present in quantum states but also gauge the potential of these states to be harnessed as quantum resources.

We investigate the limitations imposed by the mixedness of a quantum system on the degree of its imaginarity. By using the $l_{1}$ norm of imaginarity and the normalized linear entropy, we provide a trade-off relation between imaginarity and mixedness in arbitrary $d$-dimensional quantum systems, which allows us to define the ``maximally imaginary mixed state" (MIMS). For the qubit and qutrit systems, we fully identify the MIMSs, whereas for high-dimensional quantum systems, we identify a substantial class of MIMSs. Moreover, we demonstrate that this specific class of quantum states results in complementarity relations between the $1$-norm of imaginarity and the $1$-norm of mixedness, as well as between the relative entropy of imaginarity and the von Neumann entropy. Finally, we examine the evolution of the trade-off relation for single-qubit states under four distinct Markovian channels: bit flip channel, phase damping channel, depolarizing channel and amplitude damping channel.

\section{Imaginarity and mixedness}

We first recall some basic concepts related to imaginarity and mixedness of quantum systems.

\noindent{\it Resource theory of imaginarity} ~ In \cite{HG} Hickey and Gour introduced a resource theoretic framework for imaginarity. This framework is intimately connected to the resource theory of coherence, as the imaginarity is also basis dependent, with the off-diagonal elements of a density matrix exhibiting the imaginarity. Consider a $d$-dimensional Hilbert space $\mathcal{H}$ with a fixed orthonormal basis $\{|j\rangle\}_{j=0}^{d-1}$. The density matrix of a quantum state $\rho\in\mathcal{H}$ can be expressed as
\begin{equation}
\rho=\sum_{j,k}\rho_{jk}|j\rangle\langle k|.
\end{equation}
$\rho$ is termed as a real state if all the entries $\rho_{jk}$ are real. The set of all real states is denoted by $\mathcal{F}$, which defines the collection of free states within the resource theory of imaginarity. Accordingly, a density matrix featuring imaginary elements is categorized as a resource state. A quantum operation $\Lambda(\cdot)=\sum_{j}K_{j}\cdot K_{j}^{\dag}$, where $\sum_{j}K_{j}^{\dag}K_{j}=\mathbb{I}_{d}$ with $\mathbb{I}_{d}$ the $d\times d$ identity matrix, is considered free if the matrix representations of the Kraus operators $K_{j}$ are real for all $j$, that is, $\langle m|K_{j}|n\rangle\in\mathbb{R}$ for all $m,n\in[0,d-1]$, which is also referred as real operation.

A nonnegative function $M(\rho)$ is a bona fide measure of imaginarity of the state $\rho$, if it satisfies the following conditions:
(1) $M(\rho)=0$ if and only if $\rho\in\mathcal{F}$; (2) $M(\Lambda(\rho))\leq M(\rho)$ for any real operation $\Lambda$. Based on this framework, several measures of imaginarity have been proposed. In this work, we primarily utilize the $l_{1}$ norm of imaginarity defined by \cite{sc}
\begin{equation}\label{Ml1}
M_{l_{1}}(\rho)=\mathop{\mathrm{min}}_{\sigma\in\mathcal{F}}\|\rho-\sigma\|_{l_{1}}=\sum_{j\neq k}|\mathrm{Im}(\rho_{jk})|.
\end{equation}
Besides, two important measures called the $1$-norm of imaginarity \cite{HG} and the relative entropy of imaginarity \cite{qip} are also taken into account. The $1$-norm of imaginarity is given by
\begin{equation}\label{M1}
M_{1}(\rho)=\mathop{\mathrm{min}}_{\sigma\in\mathcal{F}}\|\rho-\sigma\|_{1}
=\frac{1}{2}\|\rho-\rho^{\mathrm{T}}\|_{1},
\end{equation}
where $\|A\|_{1}=\mathrm{Tr}\sqrt{A^{\dagger}A}$ denotes the trace norm of matrix $A$, and $\rho^{\mathrm{T}}$ stands for the transpose of $\rho$.
The relative entropy of imaginarity is defined as
\begin{equation}\label{Mr}
M_{r}(\rho)=\mathop{\mathrm{min}}_{\sigma\in\mathcal{F}}S(\rho\|\sigma)
=S(\mathrm{Re}(\rho))-S(\rho),
\end{equation}
where $S(\rho\|\sigma)=\mathrm{Tr}(\rho\ln\rho)-\mathrm{Tr}(\rho\ln\sigma)$ is the quantum relative entropy, $S(\rho)=-\mathrm{Tr}(\rho\ln\rho)$ is the von Neumann entropy, and $\mathrm{Re}(\rho)=\sum_{j,k}\mathrm{Re}(\rho_{jk})|j\rangle\langle k|$ is the real part of a density matrix $\rho$. It should be noted that $\mathrm{Re}(\rho)$ is a valid quantum state, as $\mathrm{Re}(\rho)=\frac{1}{2}(\rho+\rho^{\mathrm{T}})$ is a positive semi-definite matrix with unit trace.

\noindent{\it Mixedness} ~The degree of mixedness, indicative of the level of disorder within the system, can be quantified by using entropic measures related to the quantum state. The mixedness based on the normalized linear entropy is defined by
\begin{equation}\label{Sl}
S_{l}(\rho)=\frac{d}{d-1}[1-\mathrm{Tr}(\rho^{2})].
\end{equation}
It is verified that $0\leq S_{l}(\rho)\leq1$. Another measure of mixedness is the von Neumann entropy $S(\rho)$. It follows that $0\leq S(\rho)\leq\ln d$, where the maximum value of $S(\rho)$ is attained if and only if $\rho$ is the maximally mixed state.
Furthermore, the mixedness of a quantum state $\rho$ can be quantified by using the trace norm \cite{chenmf}, referred to $1$-norm of mixedness in this paper,
\begin{equation}\label{Str}
S_{1}(\rho)=1-\frac{d}{2(d-1)}\left\|\rho-\frac{1}{d}\mathbb{I}_{d}\right\|_{1},
\end{equation}
which takes values between $0$ and $1$. This measure is instrumental in deriving the trade-off relation between the $1$-norm of imaginarity and the mixedness.

\section{Trade-offs between imaginarity and mixedness and maximally imaginary mixed states }

In \cite{Singh} the authors investigated the constraints imposed on the maximum achievable quantum coherence associated with the mixedness of a system. They established a trade-off relation between the $l_{1}$-norm of coherence and the normalized linear entropy,
\begin{equation}\label{CS}
\frac{C_{l_{1}}^{2}(\rho)}{(d-1)^{2}}+S_{l}(\rho)\leq1,
\end{equation}
where $C_{l_{1}}(\rho)=\sum_{j\neq k}|\rho_{jk}|$.
Subsequently, a class of states that maximize quantum coherence for fixed mixedness is identified. These states are called the maximally coherent mixed states (MCMSs) which of the following form up to incoherent unitaries,
\begin{equation}\label{MCMS}
\rho_{m}=p|\phi_{d}\rangle\langle\phi_{d}|+\frac{1-p}{d}\mathbb{I}_{d},
\end{equation}
where $0<p\leq1$ and $|\phi_{d}\rangle=\frac{1}{\sqrt{d}}\sum_{i=0}^{d-1}|i\rangle$ is the maximally coherent state. It is evident from Eq. (\ref{Ml1}) that the $l_{1}$-norm of imaginarity does not exceed the $l_{1}$-norm of coherence, i.e., $M_{l_{1}}(\rho)\leq C_{l_{1}}(\rho)$. Thus one can easily deduce a trade-off between the $l_{1}$-norm of imaginarity and the normalized linear entropy,
\begin{equation}\label{MS}
\frac{M_{l_{1}}^{2}(\rho)}{(d-1)^{2}}+S_{l}(\rho)\leq1.
\end{equation}
This enables us to introduce the concept of maximally imaginary mixed states.

\textbf{Definition} ~ For given degree of mixedness, a $d$-dimensional quantum state is called a maximally imaginary mixed state (MIMS) if it saturates the equality in (\ref{MS}) .

The MIMS leads to a complementary relation between the $l_{1}$-norm of imaginarity and the normalized linear entropy. This relation ensures that, for a fixed level of noise measured by its mixedness, the MIMSs maximize the imaginarity. Moreover, one can see that an MIMS is also a MCMS from the fact that $M_{l_{1}}(\rho)\leq C_{l_{1}}(\rho)$. Thus the MIMSs form a significant subcategory within the MCMSs, offering substantial value for research endeavors. In the following, we explore the existence and the characterizations of MIMSs in any $d$-dimensional quantum systems.

\noindent{\it Qubit systems} ~ Let $\rho=\frac{1}{2}(I+\vec{r}\cdot\vec{\sigma})$ be a qubit state, where $\vec{r}=(r_{1},r_{2},r_{3})$ is a three-dimensional real vector with $|\vec{r}|\leq 1$ and $\vec{\sigma}=(\sigma_{1},\sigma_{2},\sigma_{3})$ given by the standard Pauli matrices. We have $M_{l_{1}}(\rho)=|r_{2}|$ and $S_{l}(\rho)=1-(r_{1}^{2}+r_{2}^{2}+r_{3}^{2})$. Thus $M_{l_{1}}^{2}(\rho)+S_{l}(\rho)=1$ if and only if $r_{1}=r_{3}=0$. In this case, for a fixed mixedness $s~(0\leq s<1)$, one gets $r_{2}=\pm\sqrt{1-s}$ and the MIMSs are of the following forms,
\begin{equation*}
\rho_{M}^{(1)}=\frac{1}{2}
\begin{pmatrix}
1  &  -\mathrm{i}\sqrt{1-s} \\
\mathrm{i}\sqrt{1-s} &   1
\end{pmatrix},~~~
\rho_{M}^{(2)}=\frac{1}{2}
\begin{pmatrix}
1  &  \mathrm{i}\sqrt{1-s} \\
-\mathrm{i}\sqrt{1-s} &   1
\end{pmatrix}.
\end{equation*}
It can be seen that $\rho_{M}^{(2)}=O\rho_{M}^{(1)}O^{\mathrm{T}}$, where $O=diag\{1,-1\}$ is a diagonal real orthogonal matrix. Hence, by setting $p=\sqrt{1-s}$ we have the following Theorem.

\textbf{Theorem 1} The MIMS for qubit system is of the following form, up to diagonal real orthogonal transformations,
\begin{equation}\label{MIMS2}
\rho_{M}=p|+\rangle\langle+|+\frac{1-p}{2}\mathbb{I}_{2},~~~0<p\leq1,
\end{equation}
where $|+\rangle=\frac{1}{\sqrt{2}}(|0\rangle+\mathrm{i}|1\rangle)$ is the maximally imaginary state.

In the context of qubits, the MIMS closely resembles the MCMS. It primarily consists of a maximally imaginary state admixed with white noise. Furthermore, it can be represented as $\rho_{M}=U\rho_{m}U^{\dag}$, where $\rho_{m}=p|\phi_{2}\rangle\langle\phi_{2}|+\frac{1-p}{2}\mathbb{I}_{2}$ with $\phi_{2}=\frac{1}{\sqrt{2}}(|0\rangle+|1\rangle)$, and $U=diag\{1,\mathrm{i}\}$ is an incoherent unitary matrix.

\noindent{\it Higher dimensional systems} ~In \cite{CHF} the authors provided a trade-off among the $l_{1}$ norm of imaginarity $M_{l_{1}}(\rho)$, linear entropy $S_{l}(\rho)$ and the normalized BZ invariant information in the presence of conjugate symmetry $I_{+}(\rho)$ for arbitrary $d$-dimensional quantum system,
\begin{equation}\label{MSI}
\frac{M_{l_{1}}^{2}(\rho)}{(d-1)^{2}}+S_{l}(\rho)+I_{+}(\rho)\leq1,
\end{equation}
where $I_{+}(\rho)=\frac{d}{d-1}\{\sum_{j,k}[\mathrm{Re}(\rho_{jk})]^{2}-\frac{1}{d}\}$. It is also showed that the equality in (\ref{MSI}) is attained if and only if the absolute values of $\mathrm{Im}(\rho_{jk})$ are all equal for $j<k$. Then $\rho$ is a MIMS if and only if $I_{+}(\rho)=0$ and $\mathrm{Im}(\rho_{jk})=:|y|$ for all $j<k$ for some number $y$.
Taking into account that $I_{+}(\rho)=0$ is equivalent to $\mathrm{Re}(\rho)=\frac{1}{d}\mathbb{I}_{d}$ \cite{Luo22}, we have the potential forms of MCMS:
\begin{equation}\label{rhoMw}
\widetilde{\rho}_{M}=
\begin{pmatrix}
\frac{1}{d}  &  \pm\mathrm{i}y  &  \cdots  &  \pm\mathrm{i}y\\
\mp\mathrm{i}y  &  \frac{1}{d}  &  \cdots  &  \pm\mathrm{i}y\\
\vdots  &  \vdots  &    &  \vdots\\
\mp\mathrm{i}y  &  \mp\mathrm{i}y  &  \cdots  &  \frac{1}{d}
\end{pmatrix}.
\end{equation}
One easily gets that $M_{l_{1}}(\widetilde{\rho}_{M})=(d^{2}-d)|y|$, and for a fixed mixedness $s$, $|y|=\frac{1}{d}\sqrt{1-s}$. Moreover, to ensure the semi-positivity of $\widetilde{\rho}_{M}$, it is necessary to specify the range of $y$. For this we need  to calculate the eigenvalues of $\widetilde{\rho}_{M}$ and ensure that the smallest one is non-negative. Next, we study the qutrit case, for which we have the following Theorem.

\textbf{Theorem 2} For a fixed mixedness $s\geq\frac{2}{3}$, the MIMS for qutrit systems has the following two forms, up to diagonal real orthogonal transformations,
\begin{equation*}
\rho_{M}^{(1)}=
\begin{pmatrix}
\frac{1}{3}  &  -\mathrm{i}y  &  -\mathrm{i}y\\
\mathrm{i}y  &  \frac{1}{3}   &  -\mathrm{i}y\\
\mathrm{i}y  &  \mathrm{i}y   &  \frac{1}{3}
\end{pmatrix},~~~
\rho_{M}^{(2)}=
\begin{pmatrix}
\frac{1}{3}   &  \mathrm{i}y  &  \mathrm{i}y\\
-\mathrm{i}y  &  \frac{1}{3}  &  \mathrm{i}y\\
-\mathrm{i}y  &  -\mathrm{i}y &  \frac{1}{3}
\end{pmatrix},
\end{equation*}
where $|y|=\frac{1}{3}\sqrt{1-s}$. For $s<\frac{2}{3}$, the maximum imaginarity in (\ref{MS}) is unattainable, that is, $M_{l_{1}}(\rho)<2\sqrt{1-s}$ for any qutrit state.

\textbf{Proof.} As shown in Eq. (\ref{rhoMw}), the potential forms of MCMS for a qutrit system are as follows:
$\rho_{M}^{(1)}$, $\rho_{M}^{(2)}$ and
\begin{equation*}
\rho_{M}^{(3)}=
\begin{pmatrix}
\frac{1}{3}           &  -\mathrm{i}y  &  \mathrm{i}y\\
\mathrm{i}y   &  \frac{1}{3}           &  \mathrm{i}y\\
-\mathrm{i}y  &  -\mathrm{i}y  &  \frac{1}{3}
\end{pmatrix},~
\rho_{M}^{(4)}=
\begin{pmatrix}
\frac{1}{3}          &  \mathrm{i}y  &  -\mathrm{i}y\\
-\mathrm{i}y &  \frac{1}{3}          &  \mathrm{i}y\\
\mathrm{i}y  &  -\mathrm{i}y &  \frac{1}{3}
\end{pmatrix},~
\rho_{M}^{(5)}=
\begin{pmatrix}
\frac{1}{3}          &  \mathrm{i}y  &  \mathrm{i}y\\
-\mathrm{i}y &  \frac{1}{3}          &  -\mathrm{i}y\\
-\mathrm{i}y &  \mathrm{i}y  &  \frac{1}{3}
\end{pmatrix},
\end{equation*}
\begin{equation*}
\rho_{M}^{(6)}=
\begin{pmatrix}
\frac{1}{3}          &  \mathrm{i}y  &  -\mathrm{i}y\\
-\mathrm{i}y &  \frac{1}{3}          &  -\mathrm{i}y\\
\mathrm{i}y  &  \mathrm{i}y  &  \frac{1}{3}
\end{pmatrix},~
\rho_{M}^{(7)}=
\begin{pmatrix}
\frac{1}{3}          &  -\mathrm{i}y  &  \mathrm{i}y\\
\mathrm{i}y  &  \frac{1}{3}           &  -\mathrm{i}y\\
-\mathrm{i}y &  \mathrm{i}y   &  \frac{1}{3}
\end{pmatrix},~
\rho_{M}^{(8)}=
\begin{pmatrix}
\frac{1}{3}         &  -\mathrm{i}y  &  -\mathrm{i}y\\
\mathrm{i}y  &  \frac{1}{3}           &  \mathrm{i}y\\
\mathrm{i}y  &  -\mathrm{i}y  &  \frac{1}{3}
\end{pmatrix}.
\end{equation*}
Let $O_{1}=diag\{1,1,-1\}$, $O_{2}=diag\{1,-1,1\}$ and $O_{3}=diag\{1,-1,-1\}$ be three diagonal real orthogonal matrices. Then we have
\begin{equation*}
\rho_{M}^{(3)}=O_{1}\rho_{M}^{(1)}O_{1}^{\mathrm{T}},~~\rho_{M}^{(4)}=O_{2}\rho_{M}^{(1)}O_{2}^{\mathrm{T}},~~
\rho_{M}^{(5)}=O_{3}\rho_{M}^{(1)}O_{3}^{\mathrm{T}},
\end{equation*}
and
\begin{equation*}
\rho_{M}^{(6)}=O_{1}\rho_{M}^{(2)}O_{1}^{\mathrm{T}},~~\rho_{M}^{(7)}=O_{2}\rho_{M}^{(2)}O_{2}^{\mathrm{T}},~~
\rho_{M}^{(8)}=O_{3}\rho_{M}^{(2)}O_{3}^{\mathrm{T}},
\end{equation*}
As $\rho_{M}^{(2)}=\rho_{M}^{(1)\mathrm{T}}$, the eigenvalues of $\rho_{M}^{(2)}$ are identical to those of $\rho_{M}^{(1)}$. Since real orthogonal transformations preserve the eigenvalues of a matrix, $\rho_{M}^{(1)}-\rho_{M}^{(8)}$ share the exact same eigenvalues. Moreover, it is direct to verify that the eigenvalues of $\rho_{M}^{(1)}$ are $\frac{1}{3}$ and $\frac{1}{3}\pm\sqrt{3}|y|$. $\rho_{M}^{(1)}$ is semi-positive definite if $|y|\leq\frac{\sqrt{3}}{9}$, or equivalently, $s\geq\frac{2}{3}$. In other words, for a fixed mixedness $s\geq\frac{2}{3}$, $\rho_{M}^{(1)}-\rho_{M}^{(8)}$ are all MIMSs, and up to diagonal real orthogonal transformations, the MIMSs are of the forms $\rho_{M}^{(1)}$ and $\rho_{M}^{(2)}$.

On the other hand, it is obvious that if $s<\frac{2}{3}$, $\rho_{M}^{(1)}-\rho_{M}^{(8)}$ are not density matrices, namely, no quantum states that saturates the equality in (\ref{MS}) for a fixed mixedness $s<\frac{2}{3}$. Therefore, for any qutrit state with $s<\frac{2}{3}$, it follows that $M_{l_{1}}(\rho)<2\sqrt{1-s}$. This completes the proof. $\Box$

It is noteworthy that $\rho_{M}^{(1)}$ and $\rho_{M}^{(2)}$, despite being MCMSs, cannot be expressed in the form of $\rho_{m}$ given in (\ref{MCMS}) by using incoherent unitaries. In fact, if $\rho_{M}^{(1)}=U\rho_{m}U^{\dag}$, where $\rho_{m}=p|\phi_{3}\rangle\langle\phi_{3}|+\frac{1-p}{3}\mathbb{I}_{3}$ and $U=diag\{e^{\mathrm{i}\theta_{1}},e^{\mathrm{i}\theta_{2}},e^{\mathrm{i}\theta_{3}}\}$ represents an incoherent unitary matrix, then the following holds true:
\begin{equation*}
\frac{p}{3}e^{\mathrm{i}(\theta_{1}-\theta_{2})}=-\mathrm{i}y,~~
\frac{p}{3}e^{\mathrm{i}(\theta_{1}-\theta_{3})}=-\mathrm{i}y,~~
\frac{p}{3}e^{\mathrm{i}(\theta_{2}-\theta_{3})}=-\mathrm{i}y.
\end{equation*}
Taking the conjugate of both sides of the second equation and multiplying it with the first and third equations, we obtain
$(\frac{p}{3})^{3}=-\mathrm{i}y^{3}$, which leads to a contradiction. The same discussions apply to the case of $\rho_{M}^{(2)}$. Hence the claim in \cite{Singh} that the MCMSs possess a unique form represented by (\ref{MCMS}) up to incoherent unitaries is not necessarily true.

For the case of general higher dimensions, from Eq. (\ref{rhoMw}) it is necessary to individually determine the eigenvalues of $\widetilde{\rho}_{M}$
for each instance among the $2^{(d^{2}-d)/2}$ cases. This undertaking is formidably difficult. Below we deal with the characterization of MIMSs for particular cases. We have the following Theorem.

\textbf{Theorem 3} For a $d$-dimensional quantum system with a fixed mixedness $s\geq\frac{2\cos(\pi/d)}{1+\cos(\pi/d)}$, the following states $O\rho_{M}^{(1)}O^{\mathrm{T}}$ and $O\rho_{M}^{(2)}O^{\mathrm{T}}$ are MIMSs, where $O$ is any diagonal real orthogonal matrix, and
\begin{equation}\label{MIMSd}
\rho_{M}^{(1)}=
\begin{pmatrix}
\frac{1}{d}  &  -\mathrm{i}y  &  \cdots  &  -\mathrm{i}y\\
\mathrm{i}y  &  \frac{1}{d}  &  \cdots  &  -\mathrm{i}y\\
\vdots  &  \vdots  &    &  \vdots\\
\mathrm{i}y  &  \mathrm{i}y  &  \cdots  &  \frac{1}{d}
\end{pmatrix},~~~
\rho_{M}^{(2)}=
\begin{pmatrix}
\frac{1}{d}  &  \mathrm{i}y  &  \cdots  &  \mathrm{i}y\\
-\mathrm{i}y  &  \frac{1}{d}  &  \cdots  &  \mathrm{i}y\\
\vdots  &  \vdots  &    &  \vdots\\
-\mathrm{i}y  &  -\mathrm{i}y  &  \cdots  &  \frac{1}{d}
\end{pmatrix}
\end{equation}
with $|y|=\frac{1}{d}\sqrt{1-s}$.

\textbf{Proof.} Since $\rho_{M}^{(2)}$ is the transpose of $\rho_{M}^{(1)}$, it suffices to compute the eigenvalues of $\rho_{M}^{(1)}$. Upon calculation, the characteristic polynomial of $\rho_{M}^{(1)}$ is given by
\begin{equation*}
\left(\lambda-\frac{1}{d}-\mathrm{i}y\right)^{d}+\left(\lambda-\frac{1}{d}+\mathrm{i}y\right)^{d}.
\end{equation*}
The $d$ distinct eigenvalues of $\rho_{M}^{(1)}$ must satisfy the following condition:
\begin{equation*}
\lambda-\frac{1}{d}-\mathrm{i}y=\left(\lambda-\frac{1}{d}+\mathrm{i}y\right)e^{\mathrm{i}\theta},
\end{equation*}
where $\theta=\frac{1}{d}\pi,\frac{3}{d}\pi,\ldots,\frac{2d-1}{d}\pi$. Thus we have
\begin{equation*}
\begin{split}
\lambda-\frac{1}{d}&=\frac{\mathrm{i}y(1+e^{\mathrm{i}\theta})}{1-e^{\mathrm{i}\theta}}\\
&=\frac{\mathrm{i}y(1+e^{\mathrm{i}\theta})(1-e^{-\mathrm{i}\theta})}{|1-e^{\mathrm{i}\theta|^{2}}}\\
&=\frac{-2y\sin\theta}{(1-\cos\theta)^{2}+\sin^{2}\theta}\\
&=-y\cot\frac{\theta}{2}.
\end{split}
\end{equation*}
If $y\geq0$, the smallest eigenvalue is
\begin{equation*}
\lambda_{min}=\frac{1}{d}-y\cot\frac{\pi}{2d}.
\end{equation*}
If $y<0$, the smallest eigenvalue is given by
\begin{equation*}
\lambda_{min}=\frac{1}{d}-y\cot\frac{(2d-1)\pi}{2d}=\frac{1}{d}+y\cot\frac{\pi}{2d}.
\end{equation*}
Hence, to guarantee that $\rho_{M}^{(1)}$ is positive semidefinite, the following inequality must be satisfied,
\begin{equation*}
\frac{1}{d}-|y|\cot\frac{\pi}{2d}\geq0,
\end{equation*}
namely,
\begin{equation*}
|y|\leq\frac{1}{d}\tan\frac{\pi}{2d}.
\end{equation*}
Taking into account that $|y|=\frac{1}{d}\sqrt{1-s}$, we get
\begin{equation*}
s\geq1-\tan^{2}\frac{\pi}{2d}=\frac{2\cos\frac{\pi}{d}}{1+\cos\frac{\pi}{d}}.
\end{equation*}
Therefore, for a fixed mixedness $s\geq\frac{2\cos(\pi/d)}{1+\cos(\pi/d)}$, $\rho_{M}^{(1)}$ is a MIMS, from which it follows that $O\rho_{M}^{(1)}O^{\mathrm{T}}$ and $O\rho_{M}^{(2)}O^{\mathrm{T}}$ are also MIMSs for any diagonal real orthogonal matrix $O$. $\Box$

It is important to note that $\rho_{M}^{(1)}$ and $\rho_{M}^{(2)}$ cannot be expressed as a mixture of the maximally imaginary state and white noise. In fact, the state $\rho_{d}=p|+\rangle\langle+|+\frac{1-p}{d}\mathbb{I}_{d}~(0<p\leq1)$ is not an MIMS when $d\geq3$, due to the fact that
\begin{equation*}
\frac{M_{l_{1}}^{2}(\rho_{d})}{(d-1)^{2}}+S_{l}(\rho_{d})=1-\frac{d(d-2)}{(d-1)^{2}}p^{2}<1, ~~~ d\geq3.
\end{equation*}

It is intriguing that the condition $\mathrm{Re}(\rho)=\frac{1}{d}\mathbb{I}_{d}$ satisfied by MIMSs remains the same for other measures of imaginarity and mixedness. For instance, if the $1$-norm of imaginarity $M_{1}(\rho)$ and the $1$-norm of mixedness $S_{1}(\rho)$ are considered, one easily derives the following trade-off relation,
\begin{equation}
\frac{d}{2(d-1)}M_{1}(\rho)+S_{1}(\rho)\leq1,
\end{equation}
with the equality holding when $\mathrm{Re}(\rho)=\frac{1}{d}\mathbb{I}_{d}$. Thus the MCMSs in (\ref{rhoMw}) satisfy the aforementioned equality and exhibit a complementarity relation between $M_{1}(\rho)$ and $S_{1}(\rho)$. Moreover, if we consider the relative entropy of imaginarity $M_{r}(\rho)$ and the von Neumann entropy $S(\rho)$ to quantify imaginarity and mixedness, respectively, we obtain
\begin{equation}
M_{r}(\rho)+S(\rho)\leq\ln d,
\end{equation}
where the equality is attained if and only if $\mathrm{Re}(\rho)=\frac{1}{d}\mathbb{I}_{d}$. Hence the MCMSs in (\ref{rhoMw}) also demonstrate a complementarity relation between $M_{r}(\rho)$ and $S(\rho)$.

\section{Evolution of the trade-off under Markovian channels }

In this section, we examine the evolution of the trade-off for single-qubit states under the influence of four distinct Markovian channels, the bit flip channel, phase damping channel, depolarizing channel and amplitude damping channel. Note that the corresponding Kraus operators for these channels are all real-valued, resulting in real operations. Thus the imaginarity of a state will not increase under the above channels. On the other hand, if we concurrently consider the mixedness of the system, it would be interesting to study the evolution of the left-hand side of the trade-off (\ref{MS}), namely, $F(\rho):=M_{l_{1}}^{2}(\rho)+S_{l}(\rho)$. Our objective is to assess whether there exists an augmented proximity of $F(\varepsilon(\rho))$ to $1$ for any non-MIMS after passing through the channel $\varepsilon$. To this end, let us consider a single-qubit state $\rho=\frac{1}{2}(I+\vec{r}\cdot\vec{\sigma})$. We have $F(\rho)=1-r_{1}^{2}-r_{3}^{2}$. Next, we compute $F(\varepsilon(\rho))$ individually for each of the previously mentioned four channels, and then compare them with respect to $F(\rho)$ to discern their relative magnitudes.

\noindent{\it Bit flip channel} ~ Let us first consider the bit flip channel which is described by the Kraus operators: $K_0^{BF}=\sqrt{p}\mathbb{I}_{2}$, $K_1^{BF}=\sqrt{1-p}\sigma_1$, where $0\leq p\leq1$. The output state is given by
\begin{equation*}
\varepsilon^{BF}(\rho)=\frac{1}{2}
\begin{pmatrix}
1+(2p-1)r_{3}  &  r_{1}-\mathrm{i}(2p-1)r_{2} \\
r_{1}+\mathrm{i}(2p-1)r_{2} &   1-(2p-1)r_{3}
\end{pmatrix}.
\end{equation*}
One can easily get
$F(\varepsilon^{BF}(\rho))=1-r_{1}^{2}-(2p-1)^{2}r_{3}^{2}$. Hence $F(\varepsilon^{BF}(\rho))\geq F(\rho)$. For a fixed $p\in(0,1)$, if $r_{3}\neq0$, then $F(\rho)$ is strictly increasing under the bit flip channel.

\noindent{\it Phase damping channel} ~ The phase damping channel is characterized by the Kraus operators: $K_0^{PD}=\sqrt{p}\mathbb{I}_{2}$, $K_1^{PD}=\sqrt{1-p}|0\rangle\langle0|$, $K_2^{PD}=\sqrt{1-p}|1\rangle\langle1|$, where $0\leq p\leq1$. Simple calculation yields that
\begin{equation*}
\varepsilon^{PD}(\rho)=\frac{1}{2}
\begin{pmatrix}
1+r_{3}  &  pr_{1}-\mathrm{i}pr_{2} \\
pr_{1}+\mathrm{i}pr_{2} &   1-r_{3}
\end{pmatrix}.
\end{equation*}
Then we have $F(\varepsilon^{PD}(\rho))=1-p^{2}r_{1}^{2}-r_{3}^{2}$. Hence we conclude that $F(\varepsilon^{PD}(\rho))\geq F(\rho)$. Moreover, for $p\neq1$, $F(\varepsilon^{PD}(\rho))$ is strictly greater than $F(\rho)$ if $r_{1}\neq0$.

\noindent{\it Depolarizing channel} ~ Quantum state after the depolarizing channel is given by $\varepsilon^{DP}(\rho)=\frac{p}{2}\mathbb{I}_{2}+(1-p)\rho$, i.e.,
\begin{equation*}
\varepsilon^{DP}(\rho)=\frac{1}{2}
\begin{pmatrix}
1+(1-p)r_{3}  &  (1-p)(r_{1}-\mathrm{i}r_{2}) \\
(1-p)(r_{1}+\mathrm{i}r_{2}) &   1-(1-p)r_{3}
\end{pmatrix}.
\end{equation*}
Thus one gets
$F(\varepsilon^{DP}(\rho))=1-(1-p)^{2}r_{1}^{2}-(1-p)^{2}r_{3}^{2}$. This implies that $F(\varepsilon^{DP}(\rho))\geq F(\rho)$. If $p\neq0$, $F(\rho)$ increases under the depolarizing channel, assuming that $r_{1}$ and $r_{3}$ are not zero simultaneously.

It can be seen that subsequent to the interaction with these three channels, the sum of $M_{l_{1}}^{2}(\rho)$ and $S_{l}(\rho)$ exhibits a strict increment for the vast majority of quantum states, provided that the parameter $p$ holds a nontrivial value. However, when considering the amplitude damping channel, a divergent trend emerges.

\noindent{\it Amplitude damping channel} ~ Now we study the evolution of $F(\rho)$ under amplitude damping channel, which can be characterized by the Kraus operators: $K_0=|0\rangle\langle0|+\sqrt{1-p}|1\rangle\langle1|$, $K_1=\sqrt{p}|0\rangle\langle1|$, where $0\leq p\leq1$. Straightforward computation shows that
\begin{equation*}
\varepsilon^{AD}(\rho)=\frac{1}{2}
\begin{pmatrix}
1+p+(1-p)r_{3}  &  \sqrt{1-p}(r_{1}-\mathrm{i}r_{2}) \\
\sqrt{1-p}(r_{1}+\mathrm{i}r_{2}) &   1-p-(1-p)r_{3}
\end{pmatrix}
\end{equation*}
and
$F(\varepsilon^{AD}(\rho))=1-(1-p)r_{1}^{2}-[p+(1-p)r_{3}]^{2}$. To compare $F(\varepsilon^{AD}(\rho))$ with $F(\rho)$, let us consider the state with a Bloch vector $\vec{r}=(\frac{\sin\theta}{\sqrt{2}},\frac{\sin\theta}{\sqrt{2}},\cos\theta)$, where $\theta\in[0,\pi]$. We have $F(\rho)=\frac{1}{2}\sin^{2}\theta$ and $F(\varepsilon^{AD}(\rho))=(1-p)[(p-\frac{1}{2})\cos^{2}\theta-2p\cos\theta+p+\frac{1}{2}]$. For $p=\frac{1}{2}$, it holds that $F(\varepsilon^{AD}(\rho))>F(\rho)$ if and only if $\theta\in(\frac{\pi}{2},\pi]$, see Fig. 1. This indicates that the relationship between $F(\varepsilon^{AD}(\rho))$ and $F(\rho)$ is state dependent. That is to say, some quantum states, after passing through the amplitude damping channel, have their sum of $M_{l_{1}}^{2}(\rho)$ and $S_{l}(\rho)$ increased, while others have it decreased, leading to a greater divergence from $1$.

\begin{figure}
\centering
\includegraphics[width=8cm]{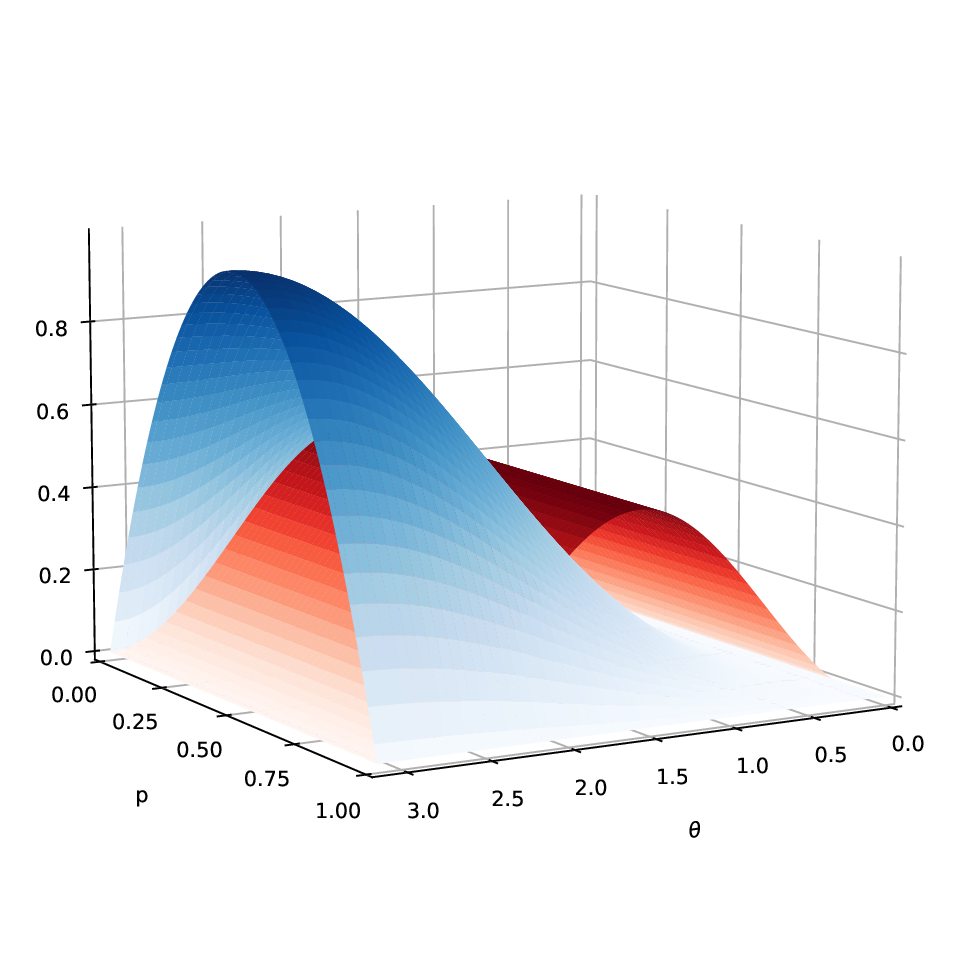}
\caption{Qubit state with Bloch vector $\vec{r}=(\frac{\sin\theta}{\sqrt{2}},\frac{\sin\theta}{\sqrt{2}},\cos\theta)$ for $\theta\in[0,\pi]$. The blue surface represents $F(\varepsilon^{AD}(\rho))$, while the red one represents $F(\rho)$. One sees that for $p=\frac{1}{2}$, $F(\varepsilon^{AD}(\rho))>F(\rho)$ if and only if $\theta\in(\frac{\pi}{2},\pi]$.}
\end{figure}

\section{Conclusion}

We have provided the trade-off relations between imaginarity and mixedness in arbitrary $d$-dimensional quantum systems, and successfully characterized the maximally imaginary mixed states (MIMSs) for both qubit and qutrit systems by utilizing the $l_{1}$ norm of imaginarity and the normalized linear entropy. Extending our investigation to higher-dimensional quantum systems, we have identified a comprehensive class of MIMSs. This specific class of quantum states not only demonstrates a complementarity relation between the $l_{1}$-norm of imaginarity and the normalized linear entropy, but also reveals the complementarity relations involving the $1$-norm of imaginarity and the $1$-norm of mixedness, as well as the relative entropy of imaginarity and the von Neumann entropy. Furthermore, we have discussed the evolution of the trade-off relation for single-qubit states under the Markovian channels bit flip, phase damping, depolarizing and amplitude damping. These channels serve as models for decoherence and dissipative processes in quantum systems. Our study elucidates how such processes affect the quantum properties of interest. Overall, our research contributes to the advancement of quantum resource theory \cite{Strel} by elucidating the correlations between imaginarity and mixedness in quantum states. It would be also interesting to investigate the constraints on other quantum resources \cite{Theu,GS,Losta,Vicen} that can be extracted from mixed quantum systems, and to identify the states that are most resourceful. Our exploration on MIMSs and the associated complementarity relations may shed new light on future researches in quantum information theory, and facilitate the understanding of imaginarity and its implications.

\vspace{2.5ex}
\noindent{\bf Acknowledgments}\, \,
This work is supported by the National Natural Science Foundation of China under Grant Nos. 11805143, 12075159 and 12171044; the
specific research fund of the Innovation Platform for Academicians of Hainan Province under Grant No. YSPTZX202215.

\end{document}